\documentclass[final,5p,times,twocolumn]{elsarticle}

\usepackage{amsmath}
\usepackage{amsthm}
\usepackage{graphicx}
\usepackage{subfigure}
\usepackage{float}
\usepackage{color}
\usepackage{amsmath }
\usepackage{amsfonts}
\usepackage{graphicx}
\usepackage{arydshln}
\usepackage{verbatim}
\usepackage{subfigure}
\usepackage{enumerate}
\usepackage{rotating}
\usepackage{threeparttable}
\usepackage{caption}
\usepackage{slashbox}
\usepackage{amsmath,xcolor}
\usepackage{graphicx}
\usepackage{epstopdf}
\usepackage{amsthm}
\usepackage{amssymb}
\usepackage{url}
\usepackage{color}
\usepackage{amsfonts}% printed Automatica style.
\biboptions{sort&compress}
\journal{Mechatronics}
\theoremstyle{remark}

\newcommand{\argmin}{\operatornamewithlimits{argmin}}
\newcommand{\argmax}{\operatornamewithlimits{argmax}}
\newcommand{\acal}{\mathcal{A}}
\newcommand{\scal}{\mathcal{S}}
\newcommand{\dcal}{\mathcal{D}}

\newcommand{\real}{\mathbb{R}}
\begin{document}

\begin{frontmatter}

\title{\huge Cloud Resource Allocation for \\ Cloud-Based Automotive Applications}

\author[AE]{Zhaojian~Li}\ead{zhaojli@umich.edu}
\author[ST]{Tianshu~Chu}\ead{cts1988@stanford.edu}
\author[AE]{Ilya~V.~Kolmanovsky}\ead{ilya@umich.edu}
\author[EECS]{Xiang~Yin\corref{cor1}}\ead{xiangyin@umich.edu}
\author[AB]{Xunyuan Yin}\ead{xunyuan@ualberta.ca}

\cortext[cor1]{The material in this paper was not presented at any IFAC conference.
Corresponding author.}

\address[AE]{Department of Aerospace Engineering, The University of Michigan, Ann Arbor, MI 48109, USA.}

\address[ST]{Department of Civil and Environmental Engineering, Stanford University, CA 94305, USA.}

\address[AB]{Department of Chemical and Materials Engineering, University of Alberta, Edmonton, AB T6G 1H9, Canada.}

\address[EECS]{Department of Electrical Engineering and Computer Science, University of Michigan, Ann Arbor, MI 48109, USA.}

\begin{abstract}
There is a rapidly growing interest in the use of cloud computing for automotive vehicles to facilitate computation and data intensive tasks. Efficient utilization of on-demand cloud resources holds a significant potential to improve future vehicle safety, comfort, and fuel economy. In the meanwhile, issues like cyber security and resource allocation pose great challenges. In this paper, we treat the resource allocation problem for cloud-based automotive systems. Both private and public cloud paradigms are considered where a private cloud provides an internal, company-owned internet service dedicated to its own vehicles while a public cloud serves all subscribed vehicles. This paper establishes comprehensive models of cloud resource provisioning for both private and public cloud-based automotive systems. Complications such as stochastic communication delays and task deadlines are explicitly considered. In particular, a centralized resource provisioning model is developed for private cloud and chance constrained optimization is exploited to utilize the cloud resources for best Quality of Services. On the other hand, a decentralized auction-based model is developed for public cloud and reinforcement learning is employed to obtain an optimal bidding policy for a ``selfish'' agent. Numerical examples are presented to illustrate the effectiveness of the developed techniques.
\end{abstract}


\end{frontmatter}

\section{Introduction}
There is growing interest in employing cloud computing in automotive applications \cite{future,li,liang2012smdp,yan2013security,whaiduzzaman2014survey,zheng2015smdp,bajcinca2015wireless,sookhak2017secure}. Ready access to distributed information and computing resources can enable computation and data intensive vehicular applications for improved safety, drivability, fuel economy, and infotainment. Several cloud-based automotive applications have been identified. For instance, a cloud-based driving speed optimizer is studied in \cite{speed} to improve fuel economy for everyday driving. In \cite{comfort}, a cloud-aided comfort-based route planner is prototyped to improve driving comfort by considering both travel time and ride comfort in route planning. A cloud-based semi-active suspension control is studied in \cite{suspension} to enhance suspension performance by utilizing road preview and powerful computation resources on the cloud.

As such, cloud computing has been both an immense opportunity and a crucial challenge for vehicular applications: opportunity because of the great potential to improve safety, comfort, and enjoyment; challenge because cyber-security and resource allocation are critical issues that need to be carefully considered. A cloud resource allocation scheme determines how a cloud server such as Amazon ``EC2'' or Google Cloud Platform distributes resources to its many clients (vehicles in our context) efficiently, effectively, and profitably. This allocation design becomes even more challenging when it comes to cloud-based automotive systems in which issues like communication delays and task deadlines arise. These complexities make a good resource allocation design a non-trivial, yet important task.

Not surprisingly, extensive studies have been dedicated to the development of efficient and profitable cloud resource allocation schemes. A dynamic bin packing method, MinTotal, is developed in \cite{bin} to minimize the total service cost. In \cite{distributed}, a  distributed and hierarchical component placement algorithm is proposed for large-scale cloud systems. A series of game theoretical cloud resource allocation approaches have also been developed, see e.g., \cite{game1,Gen-Nash,TAC,social}. However, as far as the authors are aware, a resource allocation scheme for cloud-based automotive systems that accounts for communication delays and task deadlines is still lacking.

In this paper, we develop resource allocation schemes for cloud-based automotive systems that optimally tradeoff costs and Quality of Service (QoS) with the presence of stochastic communication delays and task deadlines.  In particular, we consider allocation schemes under two cloud paradigms, private and public cloud. A private cloud is a company-owned resource center which provides computation, storage and network communication services and is only accessible by cars made by the car company. The private cloud therefore has a high level of security and information is easy and safe to share and manage. On the other hand, a public cloud relies on a third-party service provider (e.g., Amazon EC2) that provides services to all subscribed vehicles. A public cloud can eliminate the capital expenses for infrastructure acquisition and maintenance, and can provide the service on an as-needed basis.

The objectives of resource allocation are quite different between private and public cloud paradigms. Since the private cloud resources are pre-acquired, the company basically ``use them or waste them''. Therefore, the goal of private cloud resource allocation is to best utilize its resources to provide good QoS to its subscribed vehicles. Since the information exchange between vehicles and the server is more secure and convenient, the resource allocation can be achieved in a centralized manner. On the other hand, public cloud provides services to subscribed vehicles from a variety of makers, e.g., Ford, GM, Toyota, etc. Due to security and privacy issues, these vehicles typically will not share their information nor be interested in coordination; hence each vehicle becomes a ``selfish'' agent. The goal of each agent is to minimize its service cost while maintaining good QoS.

In this work, we develop mathematical models to formalize the resource allocation problems for both private and public cloud paradigms. Stochastic communication delays and onboard task deadlines are explicitly considered. A centralized resource-provisioning scheme is developed for private cloud and chance constrained optimization is employed to obtain an optimal allocation strategy. On the other hand, an auction-based bidding framework is developed for public cloud and reinforcement learning is exploited to train an optimal bidding policy  to minimize the cost while maintaining good QoS. Numerical examples are presented to demonstrate the effectiveness of the proposed schemes.

The main contributions of this paper include the following. Firstly, compared to the previous literature on cloud resource allocation, issues important to automotive vehicles such as communication delays and onboard task deadlines are explicitly treated in this paper. Secondly, resource allocation within a private cloud paradigm is formalized as a centralized resource partitioning problem. Chance constrained optimization techniques are employed to obtain the optimal partitioning by solving a convex optimization problem. Thirdly, a decentralized, auction-based bidding framework is developed for public cloud-based resource allocation and the best response dynamics assuming a constant time delay and bidding is derived. Furthermore, a Deep Deterministic Policy Gradient (DDPG) algorithm is exploited to train the optimal bidding policy with stochastic time delay and unknown bidding from other vehicles. Sensitivity analysis is also performed to show how the bidding policy can change by varying task parameters such as workload and deadline.

The rest of our paper is organized as follows. Section \ref{sec:2} describes the model of cloud resource provisioning for private cloud-based automotive systems. The problem formulation and a chance constrained optimization approach are also presented. In Section \ref{sec:3}, a numerical example is given to illustrate the allocation scheme for private cloud. The resource allocation problem with a public cloud is formalized in Section \ref{sec:4}. The best response dynamics with constant time delay and bidding is also derived. A DDPG algorithm is exploited in Section \ref{sec:5} to train the optimal bidding policy with stochastic time delay and unknown bidding from other vehicles. A numerical case study is also presented with sensitivity analysis on task parameters. Finally, conclusions are drawn in Section \ref{sec:6}.

\section{Centralized Resource Allocation with a Private Cloud}\label{sec:2}

\begin{figure}[t]
  \centering
  \includegraphics[width=0.49\textwidth]{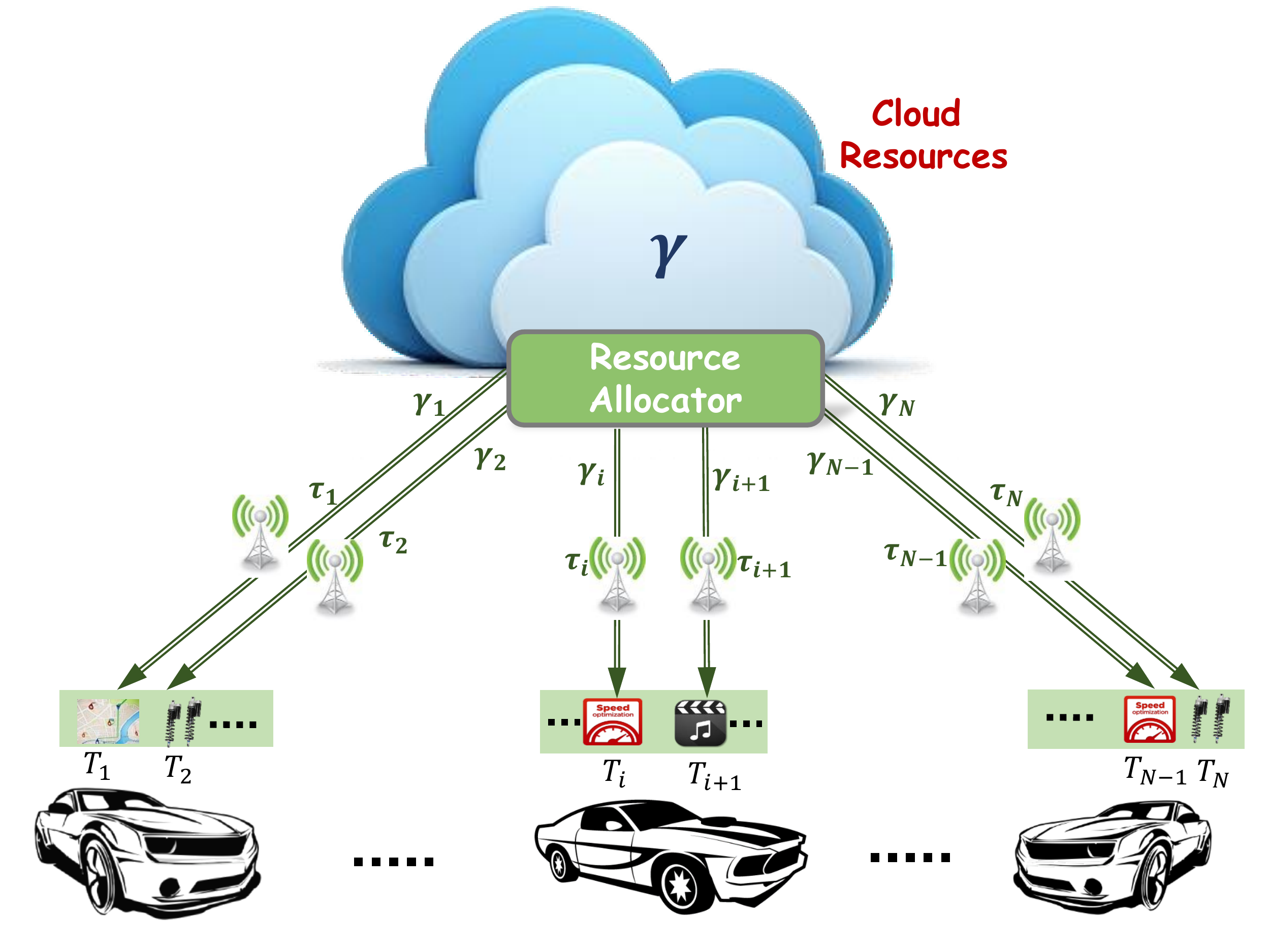}
  \caption{Schematic diagram of private cloud-based resource allocation. }\label{private}
\end{figure}

It is more secure and manageable for automotive manufacturers to acquire and maintain its own private cloud infrastructure  which provides computation, data storage and network services only to vehicles made by the manufacturer. A schematic diagram of resource allocation for private cloud-based automotive systems is illustrated in Figure~\ref{private}. Suppose that a set of cloud-based vehicular applications are available (e.g., cloud-based route planning, cloud-based suspension control, etc.) and we consider a general case that each vehicle runs a subset of these applications. Let us consider a total number of $N$ applications running on $M$ vehicles as in Figure~\ref{private}. Each application $i,\,i=1,2,\ldots,N$, corresponds to a periodic task associated with a tuple, $\mathcal{T}_i=\{T_i, w_i,d_i,\tau_i\}$, where
\begin{itemize}
\item $T_i$ is the period of task $i$ in seconds;
\item $w_i$ is the workload of task $i$ in million instructions;
\item $d_i\leq T_i$ is the deadline of task $i$ in seconds;
\item $\tau_i$ is a random time delay of the communication channel associated with task $i$ in seconds.
\end{itemize}

% and we assume $\tau_i\sim\mathcal{N}(\bar{\tau_i},\sigma_i^2)$

For each task $i$, the Quality of Service (QoS) is characterized by the following cost function adopted from \cite{zhou1}:
\begin{equation}\label{cases}
C_i(\gamma_i;\tau_i)=\begin{cases}
B_i(\frac{w_i}{\gamma_i}+\tau_i), &\mbox{if } \frac{w_i}{\gamma_i}+\tau_i\leq d_i\\
M_i, &\mbox{Otherwise},
\end{cases}
\end{equation}
where $\gamma_i$ is the process rate that the cloud resource allocator assigns to task $i$ and $\sum_{i=1}^N\,\gamma_i=\gamma$ with $\gamma$ being the total resource available on the cloud; $B_i(\cdot):\mathbb{R}^+\rightarrow \mathbb{R}^+$ is a non-decreasing function reflecting the QoS of task $i$; $M_i\geq B_i(d_i)$ is a positive scalar representing the penalty for missing the deadline; the condition $\frac{w_i}{\gamma_i}+\tau_i> d_i$ indicates that the deadline has been missed. Note that task priorities are reflected in the deadline-missing penalty $M_i$. For safety-critical tasks (e.g., cloud-based functions involved in powertrain or vehicle control), a large penalty, $M$, should be given while a small $M$ can be assigned to some non-critical tasks such as online video streaming.

Since a private cloud is a pre-acquired ``use it or waste it'' capability, the goal of resource allocation for private cloud-based automotive systems is to distribute the cloud resources to the $N$ tasks such that the total expected QoS cost as in (\ref{cases}) is minimized. Basically, the cloud collects task information (i.e., workload, deadline, time delay statistics\footnote{Note that the task period $T_i$ is not used here but we include it as one of the four task attributes for completeness. The task period will appear when it comes to the public cloud-based resource allocation in Section~\ref{sec:4}.}) of the $N$ tasks and determines how optimally to partition the total resources into $N$ parts so that the expected QoS cost is minimized. The problem can be mathematically formalized as a constrained optimization problem
\begin{equation}\label{problem1}
\begin{aligned}
&\min_{\Gamma}\, J(\Gamma)=\mathbb{E}\big[\sum_{i=1}^NC_i(\gamma_i;\tau_i)\big]\\
&\text{Subject to:}\qquad \sum_{i=1}^N\,\gamma_i=\gamma,\\
&\qquad\qquad\qquad\quad \gamma_i\geq0,\,\forall i=1,\cdots,N
\end{aligned}
\end{equation}
where $\Gamma=[\gamma_1,\gamma_2,\cdots,\gamma_N]^\text{T}$ is the vector of process rates to be optimized.

We note that the problem (\ref{problem1}) is challenging to solve due to the randomness of communication delay $\tau_i$ and the discontinuity of the cost function represented in (\ref{cases}). Motivated by the chance constrained formulation in Stochastic Model Predictive Control developments \cite{SMPC1,SMPC2,SMPC3}, we re-formulate problem (\ref{problem1}) by imposing chance constraints.
Instead of penalty for missing deadline as in (\ref{cases}), we impose chance constraints for missing deadlines of the form,
\begin{equation}\label{CC}
\Pr(\frac{w_i}{\gamma_i}+\tau_i\leq d_i)\geq 1-\alpha_i,
\end{equation}
where $\alpha_i\in (0,1)$ is a scalar representing the chance constraint of missing a deadline, $i=1,\cdots,N$. The notion of $\alpha$ can be interpreted as the upper limit of deadline missing rate specified in the QoS requirements. Applications with harsh consequences of missing a deadline can be characterized by a small $\alpha$ while larger $\alpha$ can be used for applications with mild consequences of missing a deadline.

\begin{figure}[!h]
  \centering
  \includegraphics[width=0.4\textwidth]{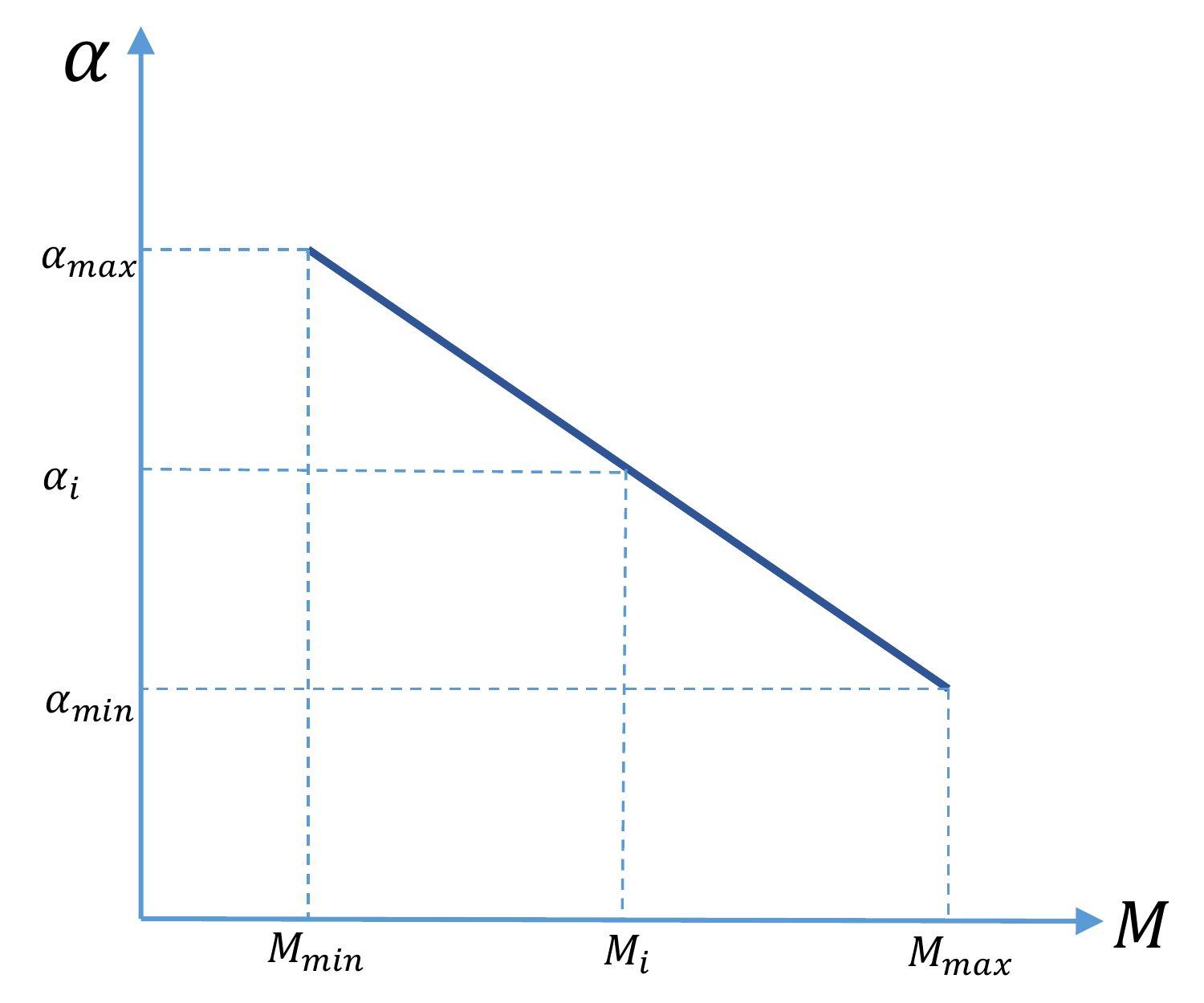}
  \caption{Linear mapping from deadline-missing penalty to chance constraint. }\label{transform}
\end{figure}

Note that the deadline-missing penalty $M$ and the chance constraint $\alpha$ are transformable. For instance, one can use the following function to map deadline-missing penalties to chance constraints:
\begin{equation}\label{convert}
  \alpha_i=\alpha_{\max}+\frac{\alpha_{\min}-\alpha_{\max}}{ M_{\max}-M_{\min}}(M_i-M_{\min}),
\end{equation}
where $\alpha_{\min}$ and $\alpha_{\max}$ are, respectively, the lower and upper bounds of the chance constraints while $M_{\min}$ and $M_{\max}$ are the corresponding lower and upper bounds of the deadline-violation penalties, respectively. These parameters need to be chosen compatibly to reflect the same QoS requirements. The example transformation in (\ref{convert}) is illustrated in Figure~\ref{transform}.

Now let us assume that the communication delays can be modeled as independent Gaussian random variables, i.e., $\tau_i\sim\mathcal{N}(\bar{\tau_i},\sigma_i^2)$. From basic probability theory, the probability of the delay taking values between $a$ and $b$,
\begin{equation}\label{inequ}
\Pr(a<\tau_i\leq b)=\frac{1}{2}\text{erf}(\frac{b-\bar{\tau_i}}{\sqrt{2}\sigma_i})-\frac{1}{2}\text{erf}(\frac{a-\bar{\tau_i}}{\sqrt{2}\sigma_i}),
\end{equation}
where $\text{erf}(x)=\frac{2}{\sqrt{\pi}}\int_0^x\;e^{-t^2}\text{d}t$ is the error function.

As a result,  from (\ref{CC}) and (\ref{inequ}), it follows that
\begin{equation}\label{equ:1}
\begin{aligned}
\Pr(\tau_i\leq d_i-\frac{w_i}{\gamma_i})=\frac{1}{2}\text{erf}(\frac{d_i-\frac{w_i}{\gamma_i}-\bar{\tau_i}}{\sqrt{2}\gamma_i})+\frac{1}{2}
\geq 1-\alpha_i.
\end{aligned}
\end{equation}

We next apply the inverse error function $\text{erf}^{-1}(\cdot)$ to both sides of (\ref{equ:1}). Since $\text{erf}^{-1}(\cdot)$ is continuous and increasing, we have
\begin{equation}\label{equ:2}
d_i-\frac{w_i}{\gamma_i}-\bar{\tau_i}\geq \sqrt{2}\sigma_i\text{erf}^{-1}(1-2\alpha_i).
\end{equation}

Note that (\ref{equ:2}) requires  the term $d_i-\bar{\tau_i}-\sqrt{2}\sigma_i\text{erf}^{-1}(1-2\alpha_i)$ to be positive so that $\gamma_i$ is feasible. This condition means that the mean of the delay $\bar{\tau_i}$ cannot be greater than the deadline $d_i$. Also, given deadline $d_i$, delay mean $\bar{\tau}_i$ and standard deviation $\sigma_i$, the minimum achievable chance constraint level, $\alpha^*$ is
\begin{equation}\label{alpha_bound}
\alpha^*=\frac{1}{2}-\frac{erf(d_i-\bar{\tau}_i)}{2\sqrt{2}\sigma_i},
\end{equation}
which defines a maximum performance bound regardless of allocated resources. For example, if $d_i=0.3$, $\bar{\tau_i}=0.2$, and $\sigma_i=0.1$, then from (\ref{alpha_bound}) we have $\alpha^*=0.3976$, which means that no matter how many resources are allocated for task $i$, the probability of missing a deadline is no less than $\alpha^*=0.3976$ due to communication delays. On the other hand, $\alpha^*<0$ means that the probability of not missing deadline can be infinitely close to 1 with enough resources.

Re-arranging terms in (\ref{equ:2}) leads to
\begin{equation}\label{equ:3}
\gamma_i\geq \rho_i=\frac{w_i}{d_i-\bar{\tau_i}-\sqrt{2}\sigma_i\text{erf}^{-1}(1-2\alpha_i)}.
\end{equation}

So far we showed that (\ref{equ:3}) and (\ref{CC}) are equivalent. Therefore, the problem in (\ref{problem1}) can be re-stated as:
\begin{equation}\label{problem2}
\begin{aligned}
&\min_{\Gamma}\, J(\Gamma)=\mathbb{E}\big[\sum_{i=1}^NB_i(\frac{w_i}{\gamma_i}+\tau_i)\big]\\
&\text{subject to:}\\
&\qquad \sum_{i=1}^N\,\gamma_i=\gamma,\\
&\qquad \gamma_i\geq \rho_i>0,\quad \forall i=1,2,\ldots,N,
\end{aligned}
\end{equation}
where $\rho_i$ are assumed to be positive and defined by (\ref{equ:3}).

Note that if we choose $B(\cdot)$ to be a convex function of $\gamma_i$, as we will show in the next section,   problem (\ref{problem2}) reduces to a convex optimization problem which can be efficiently solved with good scalability.

\section{A numerical example with a linear QoS function}\label{sec:3}
In this section, we consider a linear QoS function in the form of $B_i(\frac{w_i}{\gamma_i}+\tau_i)=b_i\cdot(\frac{w_i}{\gamma_i}+\tau_i)$ with $b_i>0$. The problem (\ref{problem2}) becomes
\begin{equation}\label{problem3}
\begin{aligned}
&\min_{\Gamma}\, J(\Gamma)=\sum_{i=1}^Nb_i\cdot(\frac{w_i}{\gamma_i}+\bar{\tau}_i)\\
&\text{subject to:}\\
&\qquad \sum_{i=1}^N\,\gamma_i=\gamma,\\
&\qquad \gamma_i\geq \rho_i>0,\quad \forall i=1,2,\ldots,N.
\end{aligned}
\end{equation}

We show that the above problem is a convex optimization problem. We first demonstrate  that the cost function $J(\Gamma)$ in (\ref{problem3}) is strictly convex in the domain $\{\Gamma=[\gamma_1,\cdots,\gamma_N]^{\text T}: \gamma_i>0, \forall i=1,2,\cdots N\}$.  Towards that end, we compute the Hessian of the cost function $J$ as
\begin{equation}
H_{xx}(J(\Gamma))=diag\{\frac{2b_1w_1}{\gamma_1^3},\cdots,\frac{2b_Nw_N}{\gamma_N^3}\},
\end{equation}
where $H_{xx}(\cdot)$ represents the Hessian matrix and $diag\{\cdot\}$ denotes the diagonal matrix with the arguments as the entries in the diagonal. Since $b_i, w_i, \gamma_i$ are positive for $i=1,2,\cdots, N$, we have $H_{xx}(J(\Gamma))$ being positive definite, which means that the cost function $J(\Gamma)$ is strictly convex. Furthermore, the constraints in (\ref{problem3}) are polytopic. Therefore, (\ref{problem3}) is a convex optimization problem that can be efficiently solved by many numerical solvers. This means even if $N$ is large, an optimal resource allocation can be efficiently computed.

We next give a numerical example with four tasks. The parameters are given in Table~\ref{tab:num} and we consider a total resource of 1, i.e., $\gamma=1$. The \textit{fmincon} function in MATLAB was exploited to solve (\ref{problem3}) and the optimized allocation strategy is,
\begin{equation}\label{allocation}
\gamma_1=0.1608,\quad \gamma_2=0.1495,\quad \gamma_3=0.3585, \quad \gamma_4=0.3312.
\end{equation}

To verify the chance constraints with the optimized allocation scheme, we run simulations under the allocation policy (\ref{allocation}) for $10^6$ times with the random delays specified in Table~\ref{tab:num}. The missing deadline violation rates for the four applications are, respectively, $0.0961, \,0.0482,\, 0.01991$, and $1.3*10^{-5}$, which are all smaller than the specified chance constraints in Table~\ref{tab:num}. This means that the specified chance constraints are satisfied under the allocation scheme (\ref{allocation}).

\begin{table*}[ht]
\caption{Parameters for numerical example}
\label{tab:num}
\centering
\begin{tabular}{|l |c|c|c|c|}
\hline
\hline
\backslashbox{Attribute}{Task} & One & Two & Three & Four\\
\hline
Workload ($w_i$, in Million instructions) & 0.02 & 0.03 & 0.1 &0.12 \\
\hline
Deadline ($d_i$, in seconds) & 0.25 & 0.35 &0.4 & 0.6\\
\hline
QoS cost scalar ($b_i$, in $\$/s$) & 1 & 2& 2 & 3\\
\hline
Delay mean ($\bar{\tau}_i$, in seconds) & 0.1 & 0.1& 0.08 & 0.11\\
\hline
Delay standard deviation ($\sigma_i$, in second) & 0.02 & 0.03& 0.02 & 0.03\\
\hline
Chance limit ($\alpha_i^*$ from (\ref{alpha_bound})) & -2.47 & -2.76& -5.67 & -5.53\\
\hline
Chance constraint ($\alpha_i$, unitless) & 0.1 & 0.05& 0.02 & 0.01\\
\hline
\hline
\end{tabular}
\end{table*}

\section{Decentralized resource allocation for public cloud paradigm}\label{sec:4}
\subsection{Problem formulation}
An automotive manufacturer may choose to subscribe its cloud-based automotive applications to a public cloud without acquiring and maintaining its own infrastructure. A public cloud such as Amazon EC2 offers an on-demand and pay-as-you-go access over a shared pool of computation resources. This public cloud provides services to a large number of automobiles from a variety of manufacturers. These vehicles may not want to share either their resource policies or task information with other vehicles, which makes it impossible to run a centralized allocation scheme as in the private cloud paradigm. Instead, each vehicle becomes a ``selfish'' client that seeks to minimize its own cost while maintaining good QoS.

\begin{figure}[t]
  \centering
  \includegraphics[width=0.49\textwidth]{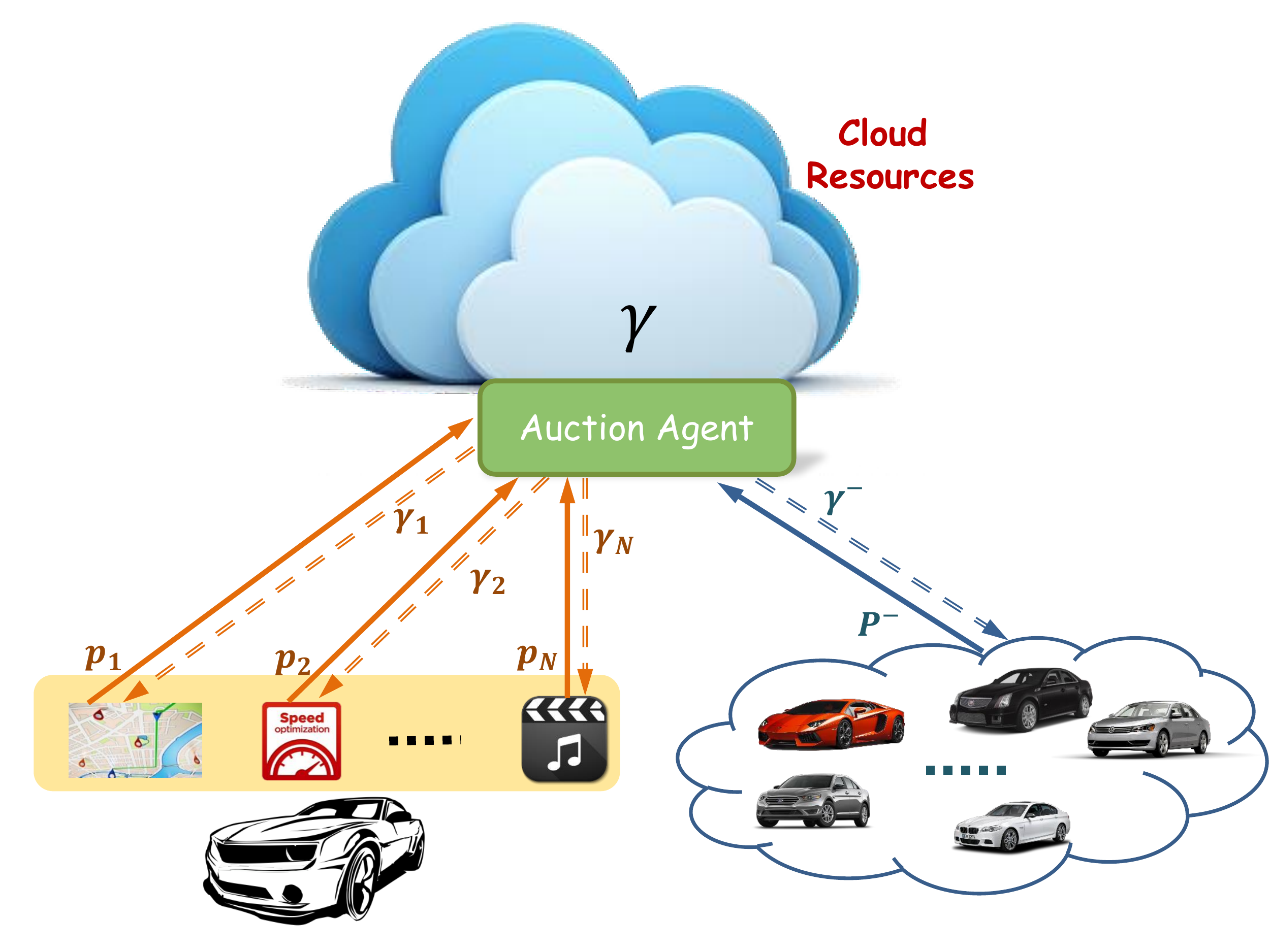}
  \caption{Schematic diagram of public cloud resource allocation }\label{public}
\end{figure}

We consider a decentralized auction-based resource allocation model as illustrated in Figure~\ref{public}. For the considered vehicle, let $N$ denote the number of tasks that are running in the vehicle and each task is associated with the same tuple $\mathcal{T}_i=\{T_i, w_i,d_i,\tau_i\}$ as defined in Section~\ref{sec:2}. The public cloud is running an auction-based resource allocation scheme, that is, each vehicular task $i,\;i=1,\cdots,N$, submits a bid $p_i$ (in US dollars per second) and obtains a proportion of the total cloud resources as:
\begin{equation}\label{proportion}
\gamma_i=\frac{p_i}{P}\cdot\gamma=\frac{p_i}{P^-+\sum_{i=1}^Np_i}\cdot \gamma,
\end{equation}
where $P$ is the sum of all bids the cloud receives from all vehicles; $P^-=P-\sum_{i=1}^Np_i$ is the cumulative bid from all \textit{other} vehicles; and $\gamma$ quantifies the total resources available on the cloud.

Since there are many other vehicles subscribed to the public cloud, it is reasonable to assume that $P^- >> \sum_{i=1}^N p_i$. From (\ref{proportion}) it follows that
\begin{equation}
\label{eq:resource}
\gamma_i \approx \frac{p_i}{P^-}\gamma,
\end{equation}
which implies that the bidding policy of the tasks can be considered independently.

\begin{figure}[!h]
  \centering
  \includegraphics[width=0.4\textwidth]{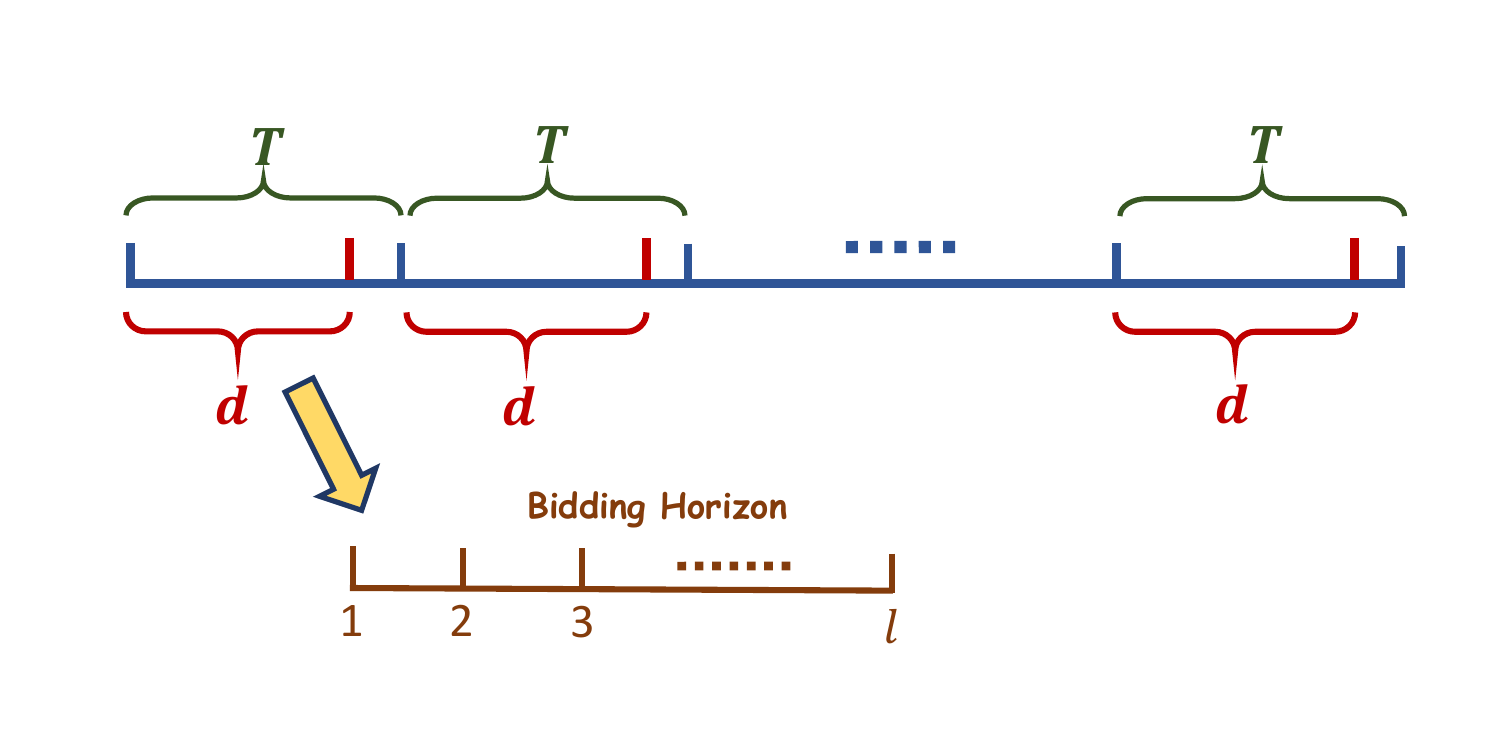}
  \caption{A general bidding model with $l$ bidding steps in one task period. }\label{bidding}
\end{figure}

We consider a general bidding model in which the time period between the beginning of a period and the deadline is composed of multiple bidding steps. As illustrated in Figure~\ref{bidding}, there are $l,\,l\geq 1$, bidding steps before the deadline in each  task period. With the QoS cost modeled in (\ref{cases}), the overall cost of task $i$ in its period $T_i$ is,
\begin{equation}\label{reward}
J_i=\sum_{t=1}^lp_{i,t}\cdot t_s+C_i(\gamma_t;\tau_i),
\end{equation}
where $p_{i,t}$ is the bidding for task $i$ at bidding step $t$; $t_s=\frac{d}{l}$ is the bidding time interval; and $C_i(\gamma_t;\tau_i)$ is the QoS cost defined in (\ref{cases}).

The goal of the vehicle is to find an optimal bidding policy to minimize the accumulated cost (\ref{reward}) for each task. We next derive the optimal bidding strategy with a preliminary assumption that $P^-$ and $\tau$ are known and constant. In Section~\ref{sec:5}, this assumption is removed.
\subsection{Best response dynamics with constant $P^-$ and $\tau$}\label{best_dynamics}
In this subsection, we seek to find the optimal bidding if bids from other vehicles ($P^-$) and the communication delays ($\tau$) are known and constant. Since $P^-$ is constant, all it matters is the total bidding. For task $i$, the optimal average bidding in the interval $[0,d_i-\tau_i]$
is defined as
\begin{equation}\label{opt_def}
\begin{aligned}
p_i^*&=\argmin_{p_i} J_i(p_i)\\
&\triangleq\argmin_{p_i} p_i\cdot (d_i-\tau_i)+C_i(p_i;\tau),
\end{aligned}
\end{equation}
where $C_i(p_i;\tau)$ is defined in (\ref{cases}) and from (\ref{eq:resource}) it follows that
\begin{equation}\label{new_C}
C_i(p_i;\tau_i)=\begin{cases}
B_i(\frac{w_iP^-}{p_i\gamma}+\tau_i), &\mbox{if } p_i\geq\frac{w_iP^-}{(d_i-\tau_i)\gamma}\\
M_i, &\mbox{Otherwise}.
\end{cases}
\end{equation}

As a result, the overall cost function $J(p_i)$ becomes
\begin{equation}\label{new_J}
J_i(p_i)=\begin{cases}
p_i\cdot (d_i-\tau_i)+B_i(\frac{w_iP^-}{p_i\gamma}+\tau_i), &\mbox{if } p_i\geq\frac{w_iP^-}{(d_i-\tau_i)\gamma}\\
p_i\cdot (d_i-\tau_i)+M_i, &\mbox{Otherwise}.
\end{cases}
\end{equation}

\begin{figure}[!h]
  \centering
  \includegraphics[width=0.45\textwidth]{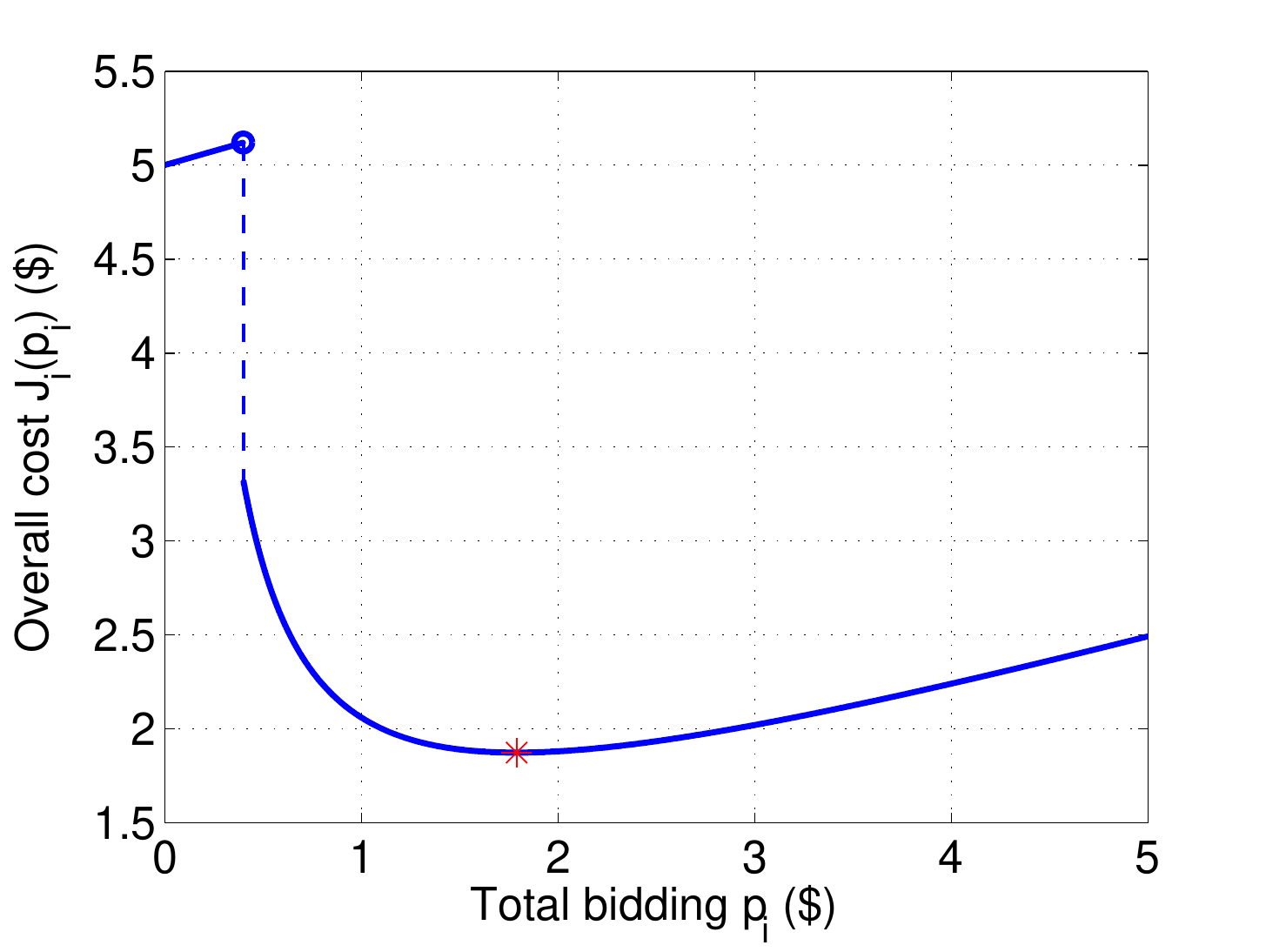}
  \caption{Overall cost as a function of total bidding with $w_i=0.06$, $d=0.4$, $\tau=0.1$, $b_i=8$, $P^-=20$, $M_i=5$, $t_s=0.05$, and $\gamma=10$. }\label{fig:min}
\end{figure}

Consider the linear QoS function $B_i(x)=b_i\cdot x$ with $b_i>0$ as in Section~\ref{sec:3}. Then there are two local minimizers in (\ref{new_J}): one associated with no bidding ($p_i^*=0$), the other corresponds to the optimal bidding with no deadline missing. The second minimizer can be represented as
\begin{equation}\label{minimizer}
p_i^*= \max\Big\{ \sqrt{\frac{b_iw_iP^-}{ (d_i-\tau_i)\gamma}}, \frac{w_iP^-}{(d_i-\tau_i)\gamma}\Big\},
\end{equation}
depending whether the minimizer of the function $p_i\cdot (d_i-\tau_i)+B_i(\frac{w_iP^-}{p_i\gamma}+\tau_i)$, i.e., $\sqrt{\frac{b_iw_iP^-}{\gamma\cdot (d_i-\tau_i)}}$, can avoid deadline missing. An example of the cost function (\ref{new_J}) with $w_i=0.06$, $d=0.4$, $\tau=0.1$, $b_i=8$, $P^-=20$, $M_i=5$, $t_s=0.05$, and $\gamma=10$ is illustrated in Figure~\ref{fig:min}. The global minimizer is $\sqrt{\frac{b_iw_iP^-}{\gamma \cdot t_s}}=1.7889$.

The assumption that the $P^-$ and $\tau$ are constant and known is unrealistic in many applications. We next exploit a reinforcement learning framework to obtain the optimal bidding policy with no assumptions on the prior knowledge of $P^-$ and $\tau$.

\section{Training optimal bidding policy using RL}\label{sec:5}
\subsection{Introduction to RL}
Reinforcement learning (RL) is a data-driven approach for adaptively evolving optimal control policies based on the real-time measurements. Unlike traditional methods, RL models the stochastic ambiguity within the framework of Markov decision processes (MDP) \cite{bellman1957markovian}, and learns the policy according to transition data observations  \cite{sutton1998reinforcement}. There are commonly three types of RL algorithms: Q-learning, policy gradient, and actor-critic. The Q-learning (or approximate Q-learning) is the traditional RL algorithm that learns a Q-function $Q_\theta(s,a)$ with model parameter $\theta$ to estimate the delayed total reward of the current state and action $a$, and performs control as $\hat{a} \in \argmax_a Q_\theta(s,a)$ \cite{szepesvari2010algorithms} based on the learned policy.  The Q-learning updates the Q-function parameters based on each observed temporal difference using stochastic gradient descent:
\begin{equation}
\theta \gets \theta + \eta \Delta_t \phi(s_t, a_t),
\end{equation}
where $\eta$ is the learning rate, $\phi(s_t,a_t)$ is the input feature vector for the learning model, and $\Delta_t$ is the sampled temporal difference with stage reward $r_t$ and discount factor $\alpha$:
\begin{equation}
\Delta_t = r_t + \alpha \max_a Q_\theta(s_{t+1},a) -Q_\theta(s_t,a_t).
\end{equation}
In the policy gradient approach,  the stochastic optimal action distribution $\pi_\theta(a|s)$ with model parameter $\theta$ is learned directly, and control action is determined as  $\hat{a} \in \argmax_a \pi_\theta(a|s)$ \cite{sutton1999policy}. Policy gradient updates the policy distribution based on each observed advantage function:
\begin{equation}
\theta \gets \theta + \eta \nabla_\theta \log\pi_\theta(a_t|s_t) (\hat{Q}^\pi(s_t,a_t)-\hat{V}(s_t)),
\end{equation}
where $\hat{Q}^\pi(s_t,a_t)$ is the sampled Q-value of $(s_t,a_t)$ by following policy $\pi_\theta$ and $\hat{V}(s_t)$ is the sampled optimal value of $s_t$.

The actor-critic approach can be regarded as a combination of both since it learns both the policy $\pi_\theta(a|s)$ and the corresponding Q-function of the policy $Q^\pi_\beta(s,a)$ \cite{konda1999actor}. The details of the actor-critic updates will be covered in the following subsection.

All the above algorithms typically assume the discrete action space which, in particular, simplifies the search for $\hat{a}$. However, in our resource allocation problem, it is more natural to consider a continuous action space since the bid should be a continuous numerical number. For this case, deterministic policy gradient algorithm was developed recently that allows to directly learn the policy $\mu(s)$ instead of the policy distribution $\pi(a|s)$, and the control is simply performed as $\hat{a} = \mu(s)$ \cite{lever2014deterministic}. Then instead of the traditional $\epsilon-$greedy exploration or Boltzmann exploration for Q-learning, we need to perform Ornstein-Uhlenbeck noise \cite{uhlenbeck1930theory} to explore with the deterministic continuous policy.

In the following sections, we first formulate the bidding-based resource allocation problem. We then propose the corresponding MDP formulation for this stochastic optimal control problem. Further, we implement a deep network based actor-critic algorithm to learn the optimal bidding strategy using deterministic policy gradient. Finally, we evaluate the performance of this algorithm using various numerical experiments.

\subsection{Training Optimal Bidding policy with deep deterministic
policy gradient}
In this section, we exploit RL to seek the optimal bidding policy. Towards that end, we first model the bidding process as a Markov Decision Process (MDP), $\mathcal{M}=\{\mathcal{S},\mathcal{A},P,r,\alpha\}$, where

\begin{itemize}
\item $\mathcal{S}=\{w_{t}, \Delta w_{t}, a_{t-1}, d_{t}\}$  represents the state space, where $w_{t}$ is the remaining workload at time $t$; $\Delta w_{t}=w_{t-1}-w_{t}$ is the recent processed work; $a_{t-1}$ represents the last bid,  and $d_{t}$ represents the remaining time until deadline. At the beginning of each period, the initial state is simply $s_{0} =[w,0, 0, d]$;
\item $\mathcal{A}\in [0,+\infty)$ is the action space representing the bidding for the task;
\item $\mathcal{P}$ is the transition matrix where each element $\mathcal{P}_{ss'}^{a}=\mathbb{P}[S_{t+1}=s'|S_{t}=s, A_{t}=a]$ is the probability that the system  transfers to $s'$ from $s$ given the current action $a$.
\item $r$ is a stage reward function given after each bidding to guide the optimal decision-making: $r(s,a)=\mathbb{E}[r_t|S_t=s, A_t=a]$.
\item $\alpha\in [0,1)$ is a discount factor that differentiates the importance of future rewards and present rewards.
\end{itemize}

Since the bidding of other vehicles is unknown so the transition matrix $\mathcal{P}$, traditional MDP optimizers such as Policy iteration and Value iteration cannot be directly applied. In this study, we exploit RL to learn an optimal bidding policy for each application. Furthermore, since the action space $\acal=[0,+\infty)$ is continuous, approximate Q-learning and stochastic policy gradient algorithms cannot be applied without action discretization. To resolve this difficulty, we exploit the
 deterministic actor-critic (DAC) algorithm. In particular, we learn both a parameterized deterministic policy function $\mu_\theta:\scal\to\acal$ to perform the bidding action, and another parameterized critic Q-function $Q_\beta:\scal\times\acal\to\real$ to evaluate the bidding strategy. The bidding and learning procedure with a typical DAC  algorithm is:
\begin{enumerate}
\item At each time step $t \leq l$, observe the state $s_{t}$.
 %so that the bid of application $i$ is restricted to 0 if it is completed: $w_{t,i}\le0$.
\item Perform a bid $\hat a_{t}$ based on the actor policy plus some random exploration perturbations, i.e., $\hat{a}_{t} = \mu_\theta(s_{t})+$perturbations.
\item Receive the cloud resource allocation $\gamma_{t}$, and update the states as
\begin{equation}
\label{eq:state_update}
\Delta w_{ t+1} =\gamma_{t}\cdot t_s, \quad w_{t+1} = w_{t}-\Delta w_{ t+1}, \quad d_{t+1} = d_{t}-t_s.
\end{equation}
\item Terminate whenever the procedure is completed: $w_{t+1} \le 0, d_{t+1} \ge 0$, or aborted: $w_{t+1} > 0, d_{t+1} = 0$.
\item Receive the current reward. If the procedure is aborted, receive a deadline-missing penalty $r_{t}=-M$; if the procedure is completed,  a cost $r_{t}=-b\cdot t$ with $b$ be a positive scalar representing the QoS cost coefficient is received; otherwise the agent receives the following state stage cost
\begin{equation}
\label{eq:reward}
r_{t} = -\hat a_{t}\cdot t_s.
\end{equation}
\end{enumerate}

In Step 2, instead of performing an $\epsilon-$greedy exploration over the entire action space, we add some Ornstein-Uhlenbeck noises into $\hat{a}_{t}$ to explore actions in the vicinity. A replay buffer is employed to store recent transitions $(s_{t},a_{t},s_{t+1},r_{t})$ so that random transitions can be sampled to train the parameterized models to reduce the effects of data correlation \cite{DQN}. When the replay buffer is filled, we can update both $\beta$ and $\theta$ in the models by exploiting $W$ randomly selected transitions from the buffer. The update of $\beta$ is similar to the one in Q-learning: First we estimate the temporal difference from each selected transition:
\begin{equation}\label{alpha}
\Delta_{t} =r_{t} + \alpha Q_\beta(s_{t+1}, \mu_\theta(s_{t+1})) - Q_\beta(s_{t},a_{t}),
\end{equation}
where $\alpha\in[0,1)$ is the discount factor. We then update the parameter in the critic Q-function using stochastic gradient descent, i.e.,
\begin{equation}
\label{eq:beta_sgd}
\beta \gets  \beta +  \frac{\eta_\beta}{W} \sum_{t=1}^W \Delta_{t} \nabla_\beta Q_\beta(s_{t}, a_{t}).
\end{equation}
Finally, we update the parameter of the actor policy function based on the Q-function estimation
\begin{equation}\label{eq:mu_sgd}
\theta \gets  \theta + \frac{ \eta_\theta}{W} \sum_{t=1}^W \nabla_\theta \mu_\theta(s_{t})\nabla_a Q_\beta(s_{t},a)|_{a=\mu_\theta(s)}.
\end{equation}
Here $\eta_\beta$ and $\eta_\theta$ are positive constants representing the learning rates. %Interestingly, Eq.~\eqref{eq:beta_sgd} and Eq.~\eqref{eq:mu_sgd} are equivalent to minibatch training where each minibatch contains $N$ i.i.d transition samples.
In this study, we apply deep neural networks as the approximation functions for both the actor and critic. This specific implementation of the deterministic actor-critic (DAC) algorithm is referred to as the deep deterministic policy gradient (DDPG) \cite{lillicrap2015continuous}. Furthermore, techniques such as experience replay \cite{DQN} and batch normalization \cite{ioffe2015batch} are also employed to improve the learning performance.
\begin{comment}
\begin{equation}
Q_\beta(s,a) = a\beta^T\nabla_\theta\mu_\theta(s) = a\beta^Ts.
\end{equation}
Then Eq.~\eqref{eq:beta_sgd} and Eq.~\eqref{eq:mu_sgd} become
\begin{equation}
\label{eq:beta_sgd1}
\beta_{t+1} =  \beta_{t} + \eta_\beta  \Delta_{i,t} a_{i,t}s_{i,t},
\end{equation}
\begin{equation}
\label{eq:mu_sgd1}
\theta_{t+1} =  \theta_{t} + \eta_\theta  (\beta_{t+1}^Ts_{i,t})s_{i,t}.
\end{equation}
\end{comment}

\begin{figure}[!h]
  \centering
  \includegraphics[width=0.47\textwidth]{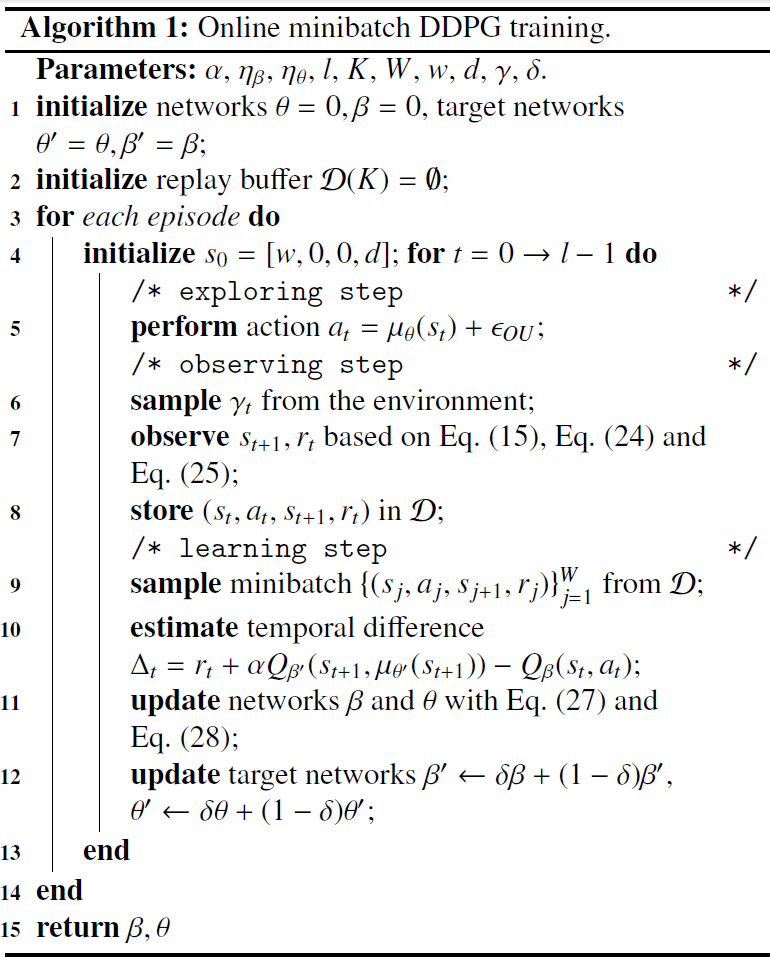}
  \caption{Algorithm 1}\label{algo:ldac}
\end{figure}

The complete DDPG algorithm is shown in Algorithm~\ref{algo:ldac}. The algorithm parameters include: the discount factor $\alpha$, learning rates $\eta_\beta$, and $\eta_\theta$ in (\ref{alpha}), (\ref{eq:beta_sgd}), and (\ref{eq:mu_sgd}), respectively; bidding horizon $l$ as in Figure~\ref{bidding}; replay buffer size $K$; mini-batch size $W, W<K$; task workload $w$; task deadline $d$; total cloud resource $\gamma$, and parameter smoothing scalar $\delta\in(0,1)$. Specifically, Line 1 initializes the network parameters and the target network parameters for smooth updating. Line 2 initializes an experience replay $\dcal$ that stores the $K$ most recent transition samples. At the beginning of each training episode (periodic task), we reset the state and sample time delay. At each time step, Line 6 performs exploration with some sampled Ornstein-Uhlenbeck noise $\epsilon_{OU}$; Line 7 samples $P^-$ from the environment; Lines 8-9 observe the system transition and add the current transition sample to the replay buffer; Lines 10-12 update the networks based on the sampled minibatch from the experience replay; Line 13 updates the corresponding target networks with a weighted sum to smooth the training. %Here we use another minibatch size $M$ instead of $N$ and revise Eq.~\eqref{eq:beta_sgd1} and Eq.~\eqref{eq:mu_sgd1} accordingly.

\subsection{Numerical experiments}

\subsubsection{Simulation setup}
\begin{figure}[!h]
  \centering
   \includegraphics[width=0.4\textwidth]{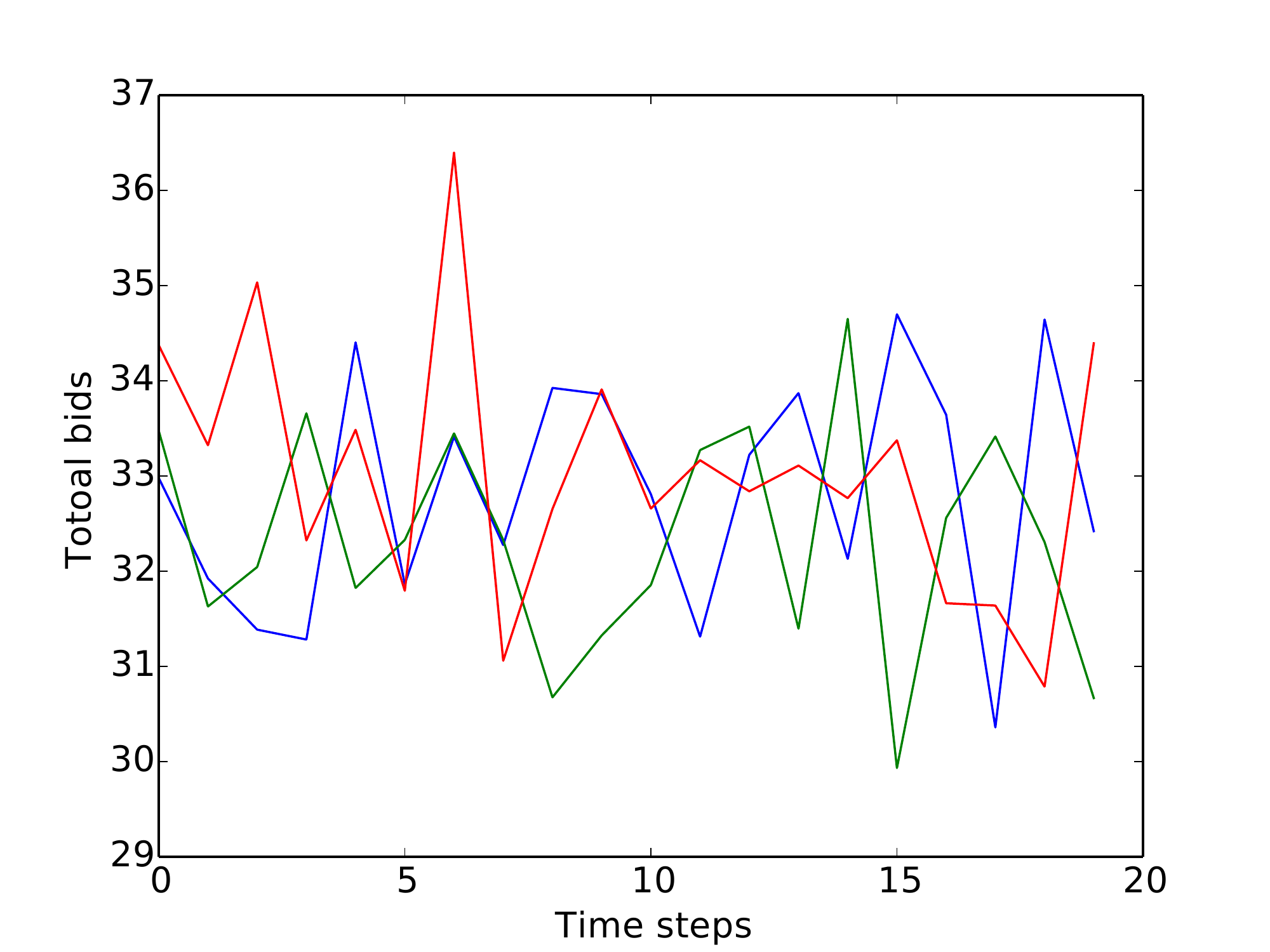}
   \caption{Three sampled $P^-$ bidding trajectories from Ornstein-Uhlenbeck process.}
    \label{fig:eg_Pminus}
\end{figure}

In this subsection, we perform simulations to illustrate the DDPG approach in Algorithm 1. Four tasks are considered in the host vehicle. The task specifications are listed in Table~\ref{tab:parameters}. The bidding policy of each application is trained separately. We define the total cloud resource as $\gamma=1$ million instructions/second. The time delays of all applications are assumed to be the same and are $\tau\sim |\mathcal{N}(0.1,0.0025)|$ in seconds. The bidding period $t_s$ is set to 0.05 seconds so $20d$ gives bidding horizon $l$ in steps and $20\tau$ gives the delay in steps. We sample $P^-$ from an Ornstein-Uhlenbeck process with $\mu=33, \theta=1, \sigma=1.5$ $\$$/second to reflect similar prices from Amazon EC2 \cite{EC2}. Three sampled trajectories of the Ornstein-Uhlenbeck process are illustrated in Figure~\ref{fig:eg_Pminus}. We set $b=2$ in the cost functions for all tasks.

For DDPG training, we train 5000 task periods with $\alpha=0.99$, $\delta=0.001$, $K=50000$, $M=32$. The actor network we use has two hidden layers with sizes 20 and 15, and the learning rate $\eta_\theta=0.00001$. We build the critic network using the same structure with learning rate is $\eta_\beta=0.0001$. We also bound the bidding at each time step as 1.5 $\$$/0.05 second to scale the output of DDPG. Our implementation is based on an open-source package DDPG\footnote{Package site: \url{https://github.com/songrotek/DDPG}.}.
%Also, we assume the maximum allowed bid is 1 $\$$ per time step.
%Also, I was considering RL to learn the variations around time, i.e., consider the peak/valley time periods. Maybe we can consider this in our next paper.
\begin{table*}[ht]
\caption{Parameters of vehicular applications for simulation.}
\label{tab:parameters}
\centering
\begin{tabular}{|c|c|c|c|c|}
\hline
\hline
\backslashbox{Attribute}{Application} & One & Two & Three & Four\\
\hline
Workload ($w$, in million instructions) & 0.02 & 0.06 & 0.1 &0.12 \\
\hline
Deadline ($d$, in seconds) & 0.5 & 0.4 &0.4 & 0.6\\
\hline
Penalty for missing deadline ($M$, in $\$$) & 2 & 2& 10 & 10\\
\hline
\hline
\end{tabular}
\end{table*}

\subsubsection{Training results}
\begin{figure}[!h]
  \centering
   \includegraphics[width=0.48\textwidth]{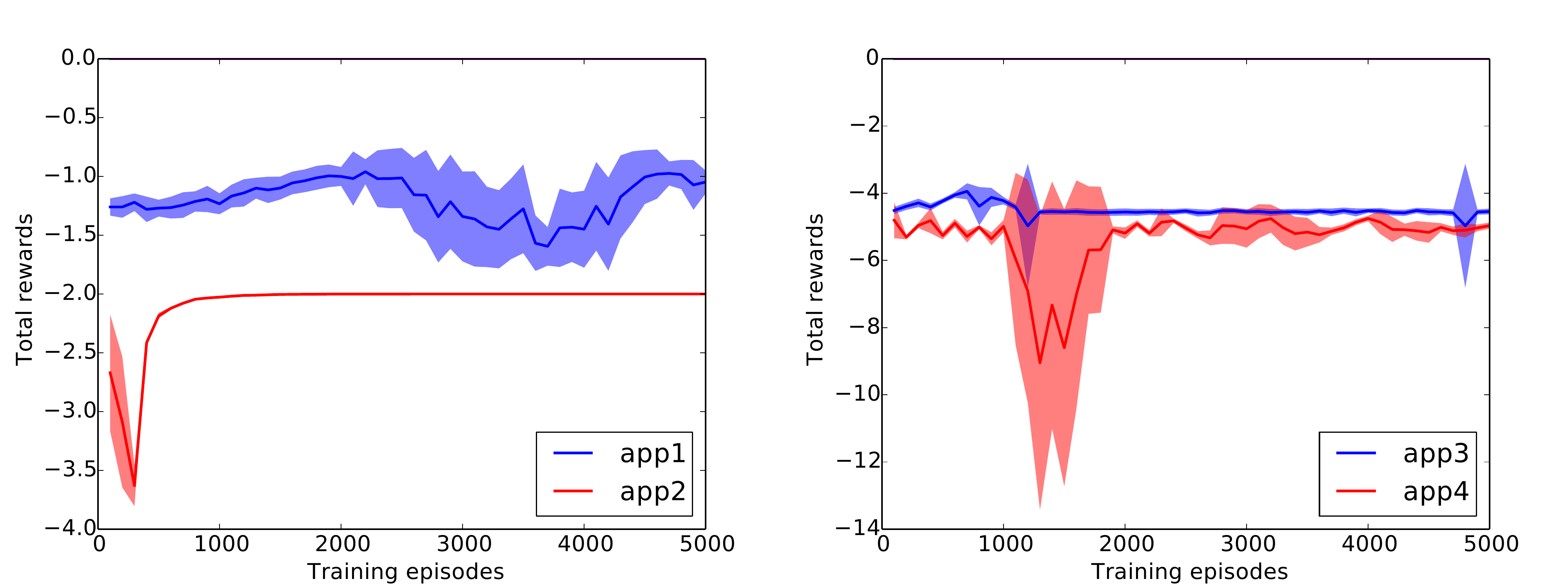}
   \caption{Total rewards vs. training episodes. Left: application 1 and 2, right: application 3 and 4.}
    \label{fig:app_train}
\end{figure}

We train the actor and critic networks with the simulation setup as described above. %Since the output of actor network must be bounded, we set the maximum allowed bid as $\$2$ per time step.
The training history of the bidding policy for the four applications is shown in Figure~\ref{fig:app_train}. The figure shows the total rewards (line is the average value and shade is the standard deviation) over 20 testing episodes from every 100 training episodes. As we can observe, the best bidding policy for application 2 is not bidding since the cost of bidding so that the deadline is not missing is more than the deadline missing penalty.

 However, the bidding policies for tasks 1, 3, and 4 do not converge. The  reason is that as we show in Section~\ref{best_dynamics}, there are two local minima in the cost function: no bidding, and minimum bidding for completing the job before deadline. Note that the second minimizer is unstable since a further small reduction on bidding may result in the convergence to the fist minimizer. So instead of using the DDPG model after the entire training, the final choice of our model is the best model during the training procedure based on the testing results as in Figure~\ref{fig:app_train}.

\begin{figure}[!h]

  \centering
   \includegraphics[width=0.48\textwidth]{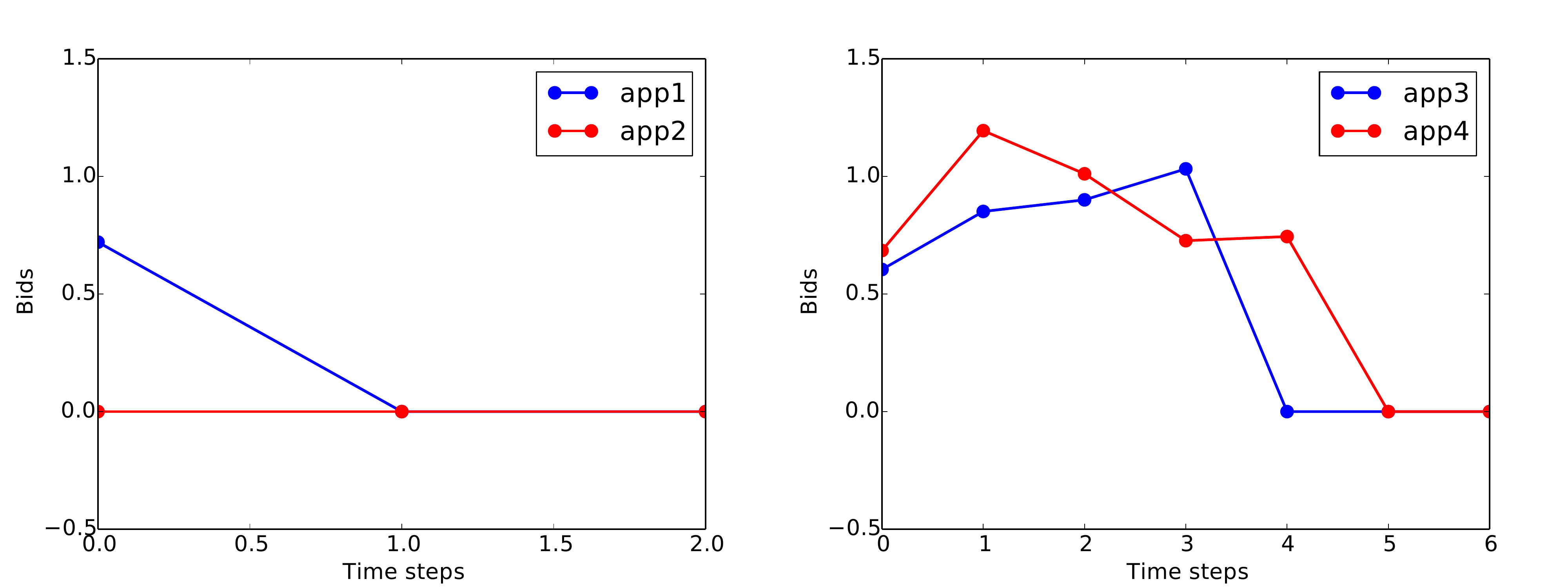}
   \caption{Trained bidding vs. time steps. Left: application 1 and 2, right: application 3 and 4.}
   \label{fig:app_bid}
\end{figure}

Next, in order to validate the optimality of our obtained policy, we investigate the trained policies with fixed $P^-=33$ $\$$/second and $\tau=0.1$ second so that we can compare with the analytical form of best bidding in Section~\ref{best_dynamics}. The results and the comparison are listed in Table~\ref{tab:policy_parameters}. We can see that DDPG almost captures either of the two local minima, depending on the amount of penalty. We also estimate the equivalent bidding rate as
\begin{equation}
\hat{p} = \frac{\sum_{t=1}^l a_t}{d-\tau},
\end{equation}
so we can compare it with the optimal bidding rate given in \eqref{minimizer}. We can see there is a small difference between $\hat{p}$ and $p^*$, which may be due to the fact that  we discretize the time horizon and round up the completion time to steps of $0.05$ second.
Figure~\ref{fig:app_bid} illustrates the detailed bidding policy at each time step for each application, where the y axis shows the actual bid per step instead of the bidding rate for easier comparison. We can see that DDPG tends to complete the process with fewer time steps but to split the bids equally among these steps.

\begin{table*}[ht]
\caption{Bidding policies for vehicular applications with fixed environment.}
\label{tab:policy_parameters}
\centering
\begin{tabular}{|c|c|c|c|c|}
\hline
\hline
\backslashbox{Variable}{Application} & One & Two & Three & Four\\
\hline
Best episode & 2200 & 4700 & 700 & 4000\\
\hline
Best average total reward & -0.96 & -2.00 & -3.94 & -4.75\\
\hline
Total bid (in $\$$) & 0.72 & 0.00 & 3.39 & 4.36\\
\hline
Equivalent bidding rate (in $\$$/second) & 1.80 & 0 & 11.30 & 8.72\\
\hline
Optimal bidding rate (in $\$$/second) & 1.83 & 0 & 11.00 & 7.92\\
\hline
Assigned workload (in million instructions) & 0.02 & 0.00 & 0.10 &0.13 \\
\hline
Completion time (in seconds) & 0.05 & NA& 0.15 & 0.20\\
\hline
\hline
\end{tabular}
\end{table*}

\subsubsection{Sensitivity analysis}
\begin{figure}[!h]
  \centering
   \includegraphics[width=0.48\textwidth]{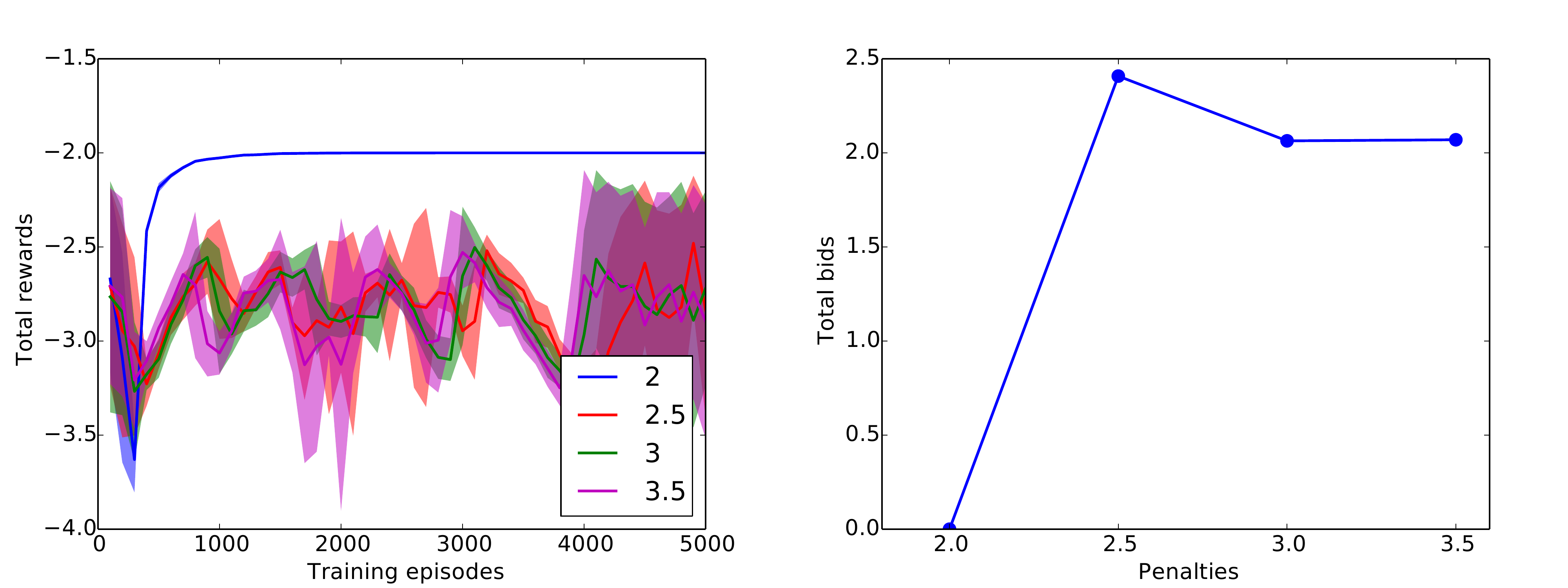}
   \caption{The impact of penalty on trained policy (in $\$$). Left: rewards during training procedure, right: total bid vs. penalty.}
   \label{fig:p_perf}
\end{figure}

In this subsection, we perform sensitivity analysis of different parameters, i.e., investigate how the bidding policy changes over task parameter variations. We choose application 2 to analyze how workload, deadline, and deadline-missing penalty can influence the obtained bidding policy. We first fix $w=0.06$ million instruction, $d=0.4$ second, and let $M=2,2.5,3,3.5$ $\$$, respectively. The results are shown in Figure~\ref{fig:p_perf}. We can see as the penalty increases, the bidding strategy switches from zero-bidding to minimum bidding for job completion. However, when $M=2.5\$$, the total rewards of these two bidding strategies are very close, so the learned policy is slightly worse than the optimal policy with total bid 2.06$\$$ and assigned workload 0.06 million instructions. When $M$ increases further, the learned policy becomes stable and optimal.

\begin{figure}[!h]
  \centering
   \includegraphics[width=0.48\textwidth]{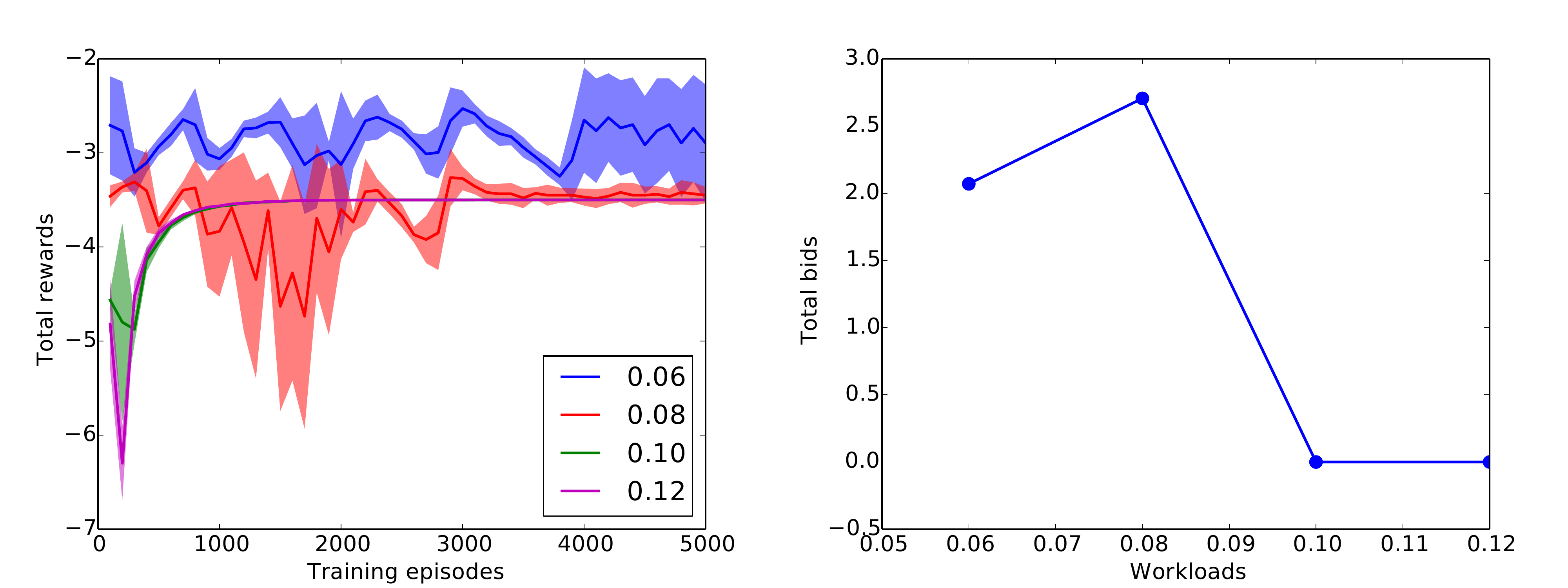}
   \caption{The impact of workload on trained policy (in million instructions). Left: rewards during training procedure, right: total bid vs. workload.}
   \label{fig:w_perf}
\end{figure}

Next, we fix $d=0.4$ second, $M=3.5$ $\$$, and change $w=0.06, 0.08, 0.1, 0.12$ million instruction, respectively. The results are shown in Figure~\ref{fig:w_perf}. We can see as the workload increases, the total bid also increases accordingly. When the total bid and QoS cost becomes larger than the penalty, the bidding strategy switches to zero-bidding.

\begin{figure}[!h]
  \centering
   \includegraphics[width=0.48\textwidth]{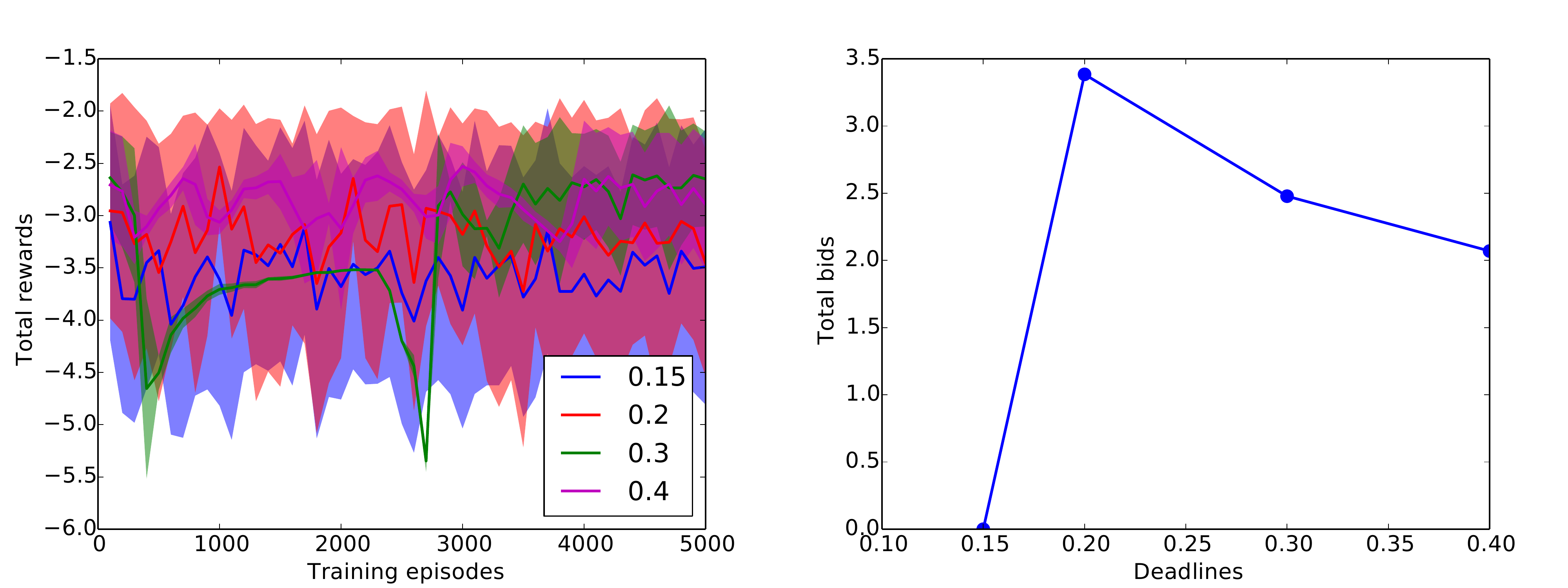}
   \caption{The impact of deadline on trained policy (in seconds). Left: rewards during training procedure, right: total bid vs. deadline.}
   \label{fig:d_perf}
\end{figure}

Finally, we fix $w=0.06$ million instruction, $M=3.5$ $\$$, and change $d=0.15, 0.2, 0.3, 0.4$ second. Similar results can be observed in Figure~\ref{fig:d_perf}: when the deadline is long enough, DDPG always bids, when the deadline is so short ($d-\tau$ may be a single step) that the application can not be completed even with the maximum bid bound, DDPG switches to zero-bidding.

\section{Conclusions}\label{sec:6}
In this paper, we studied the problem of resource allocation for cloud-based automotive systems. Resource provisioning under both private and public cloud paradigms were modeled and treated. Task deadlines and random communication delays were explicitly considered in these models. In particular, a centralized resource provisioning scheme was used to model the dynamics of private cloud provisioning and chance-constrained optimization was exploited to utilize the cloud resource to minimize the Quality of Service (QoS) cost while satisfying specified chance constraints. A decentralized auction-based bidding scheme was developed to model the public cloud resource provisioning. Best dynamics with constant bidding and constant time delays were first derived and a deep deterministic policy gradient was exploited to obtain the best bidding policy with random time delays and no prior knowledge on the random bidding from other vehicles. Numerical examples were presented to demonstrate the developed framework. We showed how the optimal bidding policy changes with parameters such as task workload and deadline.

\bibliographystyle{plain}
\bibliography{RL}

\end{document}